\documentclass[preprint,amsmath,amssymb,
 aps, prd 
nofootinbib,
nobibnotes,
]{revtex4-1}

\usepackage{graphicx}
\usepackage{dcolumn}
\usepackage{bm}
\usepackage{soul}
\usepackage{placeins}
\usepackage{mathtools,slashed}
\usepackage{hyperref}
\usepackage{color}

\begin{document}


\title{Light baryon spectra and Regge trajectories from anomalous holographic hard wall models}

\author{Rafael A. Costa-Silva}
 \email{rafaelcosta@pos.if.ufrj.br}
 \email{rc-fis@outlook.com}
\author{Henrique Boschi-Filho}%
 \email{boschi@if.ufrj.br}
 \email{hboschi@gmail.com}
\affiliation{Instituto de Física, Universidade Federal do Rio de Janeiro, 21.941-909, Rio de Janeiro, RJ, Brazil
}%

\date{\today}

\begin{abstract}
In this work we propose anomalous versions of the holographic hard wall (HW) model to describe the spectra of light baryons of spin 1/2 and 3/2, and obtain their Regge trajectories. The anomalous contributions to the dimensions of the baryonic operators of logarithm form come from a semiclassical analysis of the AdS/CFT correspondence and were used recently for glueballs and light unflavoured mesons. Inspired by these results, we first propose an anomalous dimension of the form $\Delta_{\rm anom.}=a\ln L +b$, where $a$ and $b$ are phenomenological constants to be adjusted numerically to better fit the experimental data of the PDG, and $L$ is the angular momentum of each baryonic state. Second, we discuss the case where the anomalous dimension also depends on the spin $S$ as $\Delta_{\rm anom.}=a\ln (L+S+1/2) +b$, and fix the parameters $a$ and $b$ targeting PDG data. These two models, called AHW$_1$ and AHW$_2$, give better results for light baryon masses $(M)$ in comparison with the original HW model and show approximately linear Regge trajectories $(M^2\times L)$. We also consider a third anomalous HW model in which the dimension of the baryonic operator increases as $\Delta_{\rm Lin.}=a L^c +b$, where $a$, $b$, and $c$ are constants adjusted to fit the light baryonic masses of PDG. Apart from compatible masses with PDG data, this case produces Regge trajectories that are asymptotically linear. 
\end{abstract}

\maketitle


\section{Introduction}

Understanding the nonperturbative regime of Quantum Chromodynamics (QCD) remains one of the main open problems in strong interaction physics. Although perturbative techniques successfully describe high-energy processes, the low-energy sector — where confinement and bound states emerge — still requires effective approaches. In particular, the structure of the baryon spectrum and the approximate linear behavior of Regge trajectories provide important phenomenological constraints that any realistic model should reproduce (see, {\sl e.g.}, \cite{Gross:2022hyw, Shuryak:2026pqt}). 

The gauge/gravity correspondence, originally formulated in the context of the AdS/CFT correspondence \cite{Maldacena:1997re, 
Gubser:1998bc, Witten:1998qj, Aharony:1999ti}, offers a powerful framework to investigate strongly coupled gauge theories through dual gravitational descriptions. Although QCD is neither supersymmetric nor conformal, bottom-up holographic constructions have proven useful in incorporating essential features of low-energy hadron physics in an effective manner (see, {\sl e.g.} \cite{Ammon:2015wua, Zaanen:2015oix}).

Among these constructions, the hard wall (HW) model \cite{Polchinski:2001tt,Polchinski:2002jw, Boschi-Filho:2002xih, Boschi-Filho:2002wdj} introduces an infrared cutoff in AdS space to mimic confinement, generating discrete mass spectra for hadronic states. This framework has been successfully applied to glueballs, mesons, and  baryons
\cite{Boschi-Filho:2002xih, Boschi-Filho:2002wdj, deTeramond:2005su, Erlich:2005qh, Boschi-Filho:2005xct}. Despite these successes, some limitations are well known. In particular, the original Hard Wall setup does not naturally produce asymptotically linear Regge trajectories, and the degeneracy patterns depend sensitively on the choice of scaling dimensions assigned to the boundary operators. Linear Regge trajectories were obtained soon after with the soft wall model \cite{Karch:2006pv, Colangelo:2008us}. 


The hard and soft wall models were applied to a variety of problems, from spectra and Regge trajectories \cite{Forkel:2007ru, Vega:2008af, Afonin:2009pd, Wang:2009wx, Kim:2009bp, Alvares:2011wb, FolcoCapossoli:2013eao, 
Gorsky:2013dda, Harada:2014lza, Capossoli:2021ope},
the determination of form factors,  decay constants and couplings \cite{Grigoryan:2007my, Kwee:2007dd, Grigoryan:2007wn, Kwee:2007nq, JeffersonLab:2008jve, Colangelo:2012ipa, Mamedov:2016ype, Ballon-Bayona:2017bwk, MartinContreras:2019kah, Momeni:2020bmy},  high energy scattering \cite{BallonBayona:2007qr, BallonBayona:2007rs, Bayona:2010hb, Nishio:2011xa, Nishio:2011xz, Brower:2012mk, Bolognesi:2012pi, Watanabe:2012uc, Costa:2013uia, Watanabe:2013spa, Agozzino:2014qta, Dehghani:2015jva, FolcoCapossoli:2015hub, Kovensky:2018xxa, 
Xie:2019soz, Mamo:2021cle, Ghodrati:2023uef, Armoni:2025dfq}, 
the holographic construction of the Cornell potential \cite{Boschi-Filho:2005nmp, Andreev:2006ct, White:2007tu, Bruni:2018dqm}, dynamical corrections to the soft wall model \cite{Gursoy:2007cb, Gursoy:2007er, dePaula:2008fp, Li:2013oda, FolcoCapossoli:2015jnm, FolcoCapossoli:2016fzj, Rodrigues:2016kez, Ballon-Bayona:2020qpq}, finite temperature and density effects and the confinement-deconfinement phase transition \cite{Andreev:2006eh, Boschi-Filho:2006hfm, Herzog:2006ra, BallonBayona:2007vp, Cai:2007zw, Gursoy:2009zza, Miranda:2009uw, Mamani:2013ssa, Mamani:2018uxf, Braga:2020opg, Colangelo:2009ra, Jo:2009xr, Craps:2013iaa, Rinaldi:2021xat, Lee:2021rag, Singh:2022obu, Rinaldi:2023iwf, Chen:2023bms, Wang:2024szr, Braga:2025eiz, Nasibova:2025wnw, Fujii:2025umi, Bartolini:2022rkl, Braga:2022yfe}, magnetic susceptibility, magnetic and inverse magnetic catalysis \cite{Gorsky:2009ma, Bu:2012mq, Mamo:2015dea, Dudal:2015wfn, Rodrigues:2017cha, Rodrigues:2017iqi, Rodrigues:2018pep, Rodrigues:2018chh}, entanglement \cite{Barbon:2008sr, Grieninger:2023ehb}, weak-intaracting holographic QCD \cite{Gazit:2008gz}, technicolor modeling \cite{Dietrich:2008up, Dietrich:2009af, Kutasov:2012uq}, the muon $g-2$ problem \cite{Hong:2009zw, Leutgeb:2019gbz, Colangelo:2024xfh, Mager:2025pvz, Shen:2026bko}, gluon and pion condensates  \cite{Cappiello:2010egv, Albrecht:2010eg, Chen:2024cxh}, $1/N_c$ corrections and large $N$ effects \cite{Lee:2010dh, Li:2023qzs},  nuclei properties \cite{Kim:2011ey, Park:2011zp, Agozzino:2013zgy, Park:2016hvb, MartinContreras:2022lxl}, and to fluids and condensed matter systems \cite{Puletti:2011pr, Blake:2012tp, Kulkarni:2012in, Swingle:2013rda, Fujita:2014mqa, Argurio:2017irz, Wang:2020isw}. 


Despite these many successes, the non-linearity of the Regge trajectories obtainable from the hard wall model persists. This situation changed by considering anomalous dimensions for the hadronic operators in the case of glueballs \cite{Costa-Silva:2023vuu} and light unflavoured mesons 
\cite{Costa-Silva:2024dlq}. 

Experimental data compiled by the Particle Data Group (PDG) \cite{PDG:2022pth} show a rich structure for nucleon and delta resonances, including approximate linear Regge behavior and systematic organization of parity. Reproducing these features within a holographic framework requires a careful treatment of the relation between bulk fermion masses and the scaling dimensions of the corresponding boundary operators. In interacting quantum field theories, operator dimensions generally receive anomalous contributions due to renormalization effects (see, {\sl e.g.} \cite{Gross:2022hyw, Shuryak:2026pqt}). 

Motivated by semiclassical analyses of large-spin operators in the gauge/gravity correspondence \cite{Gubser:2002tv} and by previous successful applications of anomalous dimensions in holographic models for glueballs and light mesons \cite{Costa-Silva:2023vuu,Costa-Silva:2024dlq}, we investigate the impact of including anomalous contributions in the baryonic hard wall framework.

In this work, we first revisit the description of fermionic fields in AdS$_5$, deriving the equations of motion for the spin-1/2 and spin-3/2 fields and reviewing the relation between the bulk mass parameters and the operator dimensions. We reconstruct the baryon spectrum in the original Hard Wall model, allowing for distinct canonical dimensions for different spins. We then introduce logarithmic realizations of anomalous dimensions and analyze their phenomenological consequences through a numerical minimization procedure based on experimental data. In addition, we explore a linear variant of the anomalous dimension aimed at improving the asymptotic behavior of Regge trajectories.

Our main objective is to assess whether the inclusion of anomalous dimensions within the Hard Wall approach can provide a more realistic description of light baryon spectra while preserving the simplicity of the bottom-up holographic construction.

The paper is organized as follows. In Section II we review fermions in AdS space. Section III presents the experimental input and statistical definitions. In Section IV we discuss the original and anomalous HW models. The numerical results are discussed in Section V, followed by an anomalous linear hard wall model in Section VI, and we conclude in Section VII with a discussion of the results and possible future developments.


\FloatBarrier
\newpage

\section{PDG data and Statistical definitions}

\begin{table}
\centering
\begin{tabular}{|c|c|c|c|c|}
\hline
\hline
\, $L$ \, & \, $S$ \,  & Nucleon Resonances \\ \hline
 $0$   & $1/2$ & $N_{1/2+}(939)$\\
 
 $1$  & $1/2$&$N_{1/2-}(1535),\quad N_{3/2-}(1520)$ \\
 
 $1$ & $3/2$ &$N_{1/2-}(1650),\quad N_{3/2-}(1700),\quad N_{5/2-}(1675)$\\
 
 $2$ & $1/2$ & $N_{3/2+}(1720),\quad N_{5/2+}(1680)$  \\
 
 $2$ & $3/2$ & $N_{1/2+}(1880),\quad N_{3/2+}(1900),\quad N_{5/2+}(1870),\quad N_{7/2+}(1990)$ \\
 
 $3$ & $1/2$ & $N_{5/2-}(2200),\quad N_{7/2-}(2190)$ \\
 $3$ & $3/2$ & $N_{9/2-}(2250)$ \\
 $4$ & $1/2$ & $N_{9/2+}(2220)$ \\
 $5$ & $1/2$ & $N_{11/2-}(2600)$ \\
 $6$ & $1/2$ & $N_{13/2+}(2700)$ \\
 \hline\hline
\end{tabular}
\caption{\label{pdgN} Nucleon resonances states $N_{J,P}(M)$ with masses $M$ in MeV from PDG.}
\end{table}

\vskip 0.5cm
\begin{table}
\centering
\begin{tabular}{|c|c|c|c|c|}
\hline
\hline
\, $L$ \, & \, $S$ \, & Delta Resonances \\ \hline
 $0$   & $3/2$ & $\Delta_{3/2+}(1232)$ \\
 
 $1$  & $1/2$&$\Delta_{1/2-}(1620),\quad \Delta_{3/2-}(1700)$\\
 
 $2$ & $3/2$ &$\Delta_{1/2+}(1910),\quad \Delta_{3/2+}(1920),\quad \Delta_{5/2+}(1905),\quad \Delta_{7/2+}(1950)$\\
 
 $3$ & $1/2$ & $\Delta_{7/2-}(2200)$\\
 
 $4$ & $3/2$ & $\Delta_{7/2+}(2390),\quad\Delta_{9/2+}(2300),\quad\Delta_{11/2+}(2420)$\\
 $6$ & $3/2$ & $\Delta_{15/2+}(2950)$\\
 
 \hline\hline
\end{tabular}
\caption{\label{pdgDelta} Delta resonances states $\Delta_{J,P}(M)$ with masses $M$ in MeV from PDG.}
\end{table}

   In this section, we present the masses for the PDG nucleon and delta resonances \cite{PDG:2022pth} and the  statistical functions used to perform the numerical analysis. The resonances are separated into their respective families for each fixed $L$ and $S$, where these correspond, respectively, to the orbital angular momentum and the intrinsic spin, in Tables \ref{pdgN} and \ref{pdgDelta}. Note that these states are also classified by their parity $P$, which is determined by their angular momentum states: $L$ even (odd) corresponds to positive (negative) parity.

In particular, we begin by presenting the functions that we use to determine the best fit of PDG masses from the proposed HW models. We first define the deviation between our estimated  $m_i$ and the PDG  $M_i$ masses 
   \begin{equation}
        \delta_{i}=\frac{|M_i-m_i|}{M_i}.
    \end{equation}
   In addition, we define the function $Q$,
\begin{equation}\label{qfunction}
Q=\sqrt{\sum_{i}\delta_i^2}, 
   \end{equation}
   which will be minimized to determine the best fit of our models compared to PDG data. 



\section{Brief review of fermions in AdS space}\label{Review:fermions}

Here, in this section, we briefly review the basic properties of the AdS space and how to describe fermionic fields with spins 1/2 and 3/2 in its bulk. The AdS$_5$ space can be conveniently defined by the Poincaré patch with metric \cite{Aharony:1999ti}
\begin{equation}\label{metric}
        ds^2=g_{MN}dx^Mdx^N
        = \frac{R^2}{z^2}
        \left(\eta^{\mu\nu}dx_\mu^2dx_\nu^2+dz^2\right), 
    \end{equation}
where $\eta_{\mu\nu}= \rm diag (-1,1,1,1)$ is the flat four-dimensional Minkowski metric with $\mu,\nu=0,1,2,3$ while $M,N=0,1,2,3,5$ tags the five-dimensional AdS$_5$ space. The holographic coordinate $z$ is defined in the interval $[0,\infty)$, and $z=0$ corresponds to a boundary of the AdS space. The fields defined in the bulk of the AdS$_5$ space typically span over the five coordinates $x^M$.


\subsection{Fermions with spin $1/2$}    

    The Dirac equation in the 4-dimensional Minkowski space is
    \begin{equation}\label{minkDE}
        (\gamma^{\mu}\partial_{\mu}-m)\psi(x)=0,
    \end{equation}
    with,
    \begin{equation}\label{minkgamma}
        \{\gamma^{\mu},\gamma^{\nu}\}=2\eta^{\mu\nu}.
    \end{equation}

    Defining the Dirac equation in a 5-dimensional flat spacetime requires five gamma matrices that satisfy an extension of the relation  (\ref{minkgamma}). A natural choice is obtained by making \cite{Henningson:1998cd, Mueck:1998iz}
    \begin{equation}
        \Gamma_{\mu}=\gamma_{\mu}\quad;\quad\Gamma_{5}=\gamma_{5},
    \end{equation}
    where $\gamma_{5}=-i\gamma^{0}\gamma^{1}\gamma^{2}\gamma^{3}$. The matrices defined in this way satisfy the relation
    \begin{equation}
        \{\Gamma_{M},\Gamma_{N}\}=2\eta_{MN}.
    \end{equation}

    If we want to obtain the Dirac equation in AdS$_5$, we need to use vielbein. Locally the curved spacetime at a point P can have the metric tensor given by the Minkowski metric in a special coordinate system. Let $\xi^{\alpha}$ be the coordinates of this point in this frame. The vielbein is a set of vectors given by
    \begin{equation}
        e_{\mu}^{\alpha}=\partial_{\mu}\xi^{\alpha},
    \end{equation}
    where $x$, present in the derivative, are the coordinates of P in the general frame with metric $g_{\mu\nu}(x)$. The following relation holds:
    \begin{equation}
        \eta_{\alpha\beta}e^{\alpha}_{\mu}e^{\beta}_{\nu}=g_{\mu\nu}.
    \end{equation}

    Generally, if we define
    \begin{equation}
        \tilde{\Gamma}_{M}=e^{N}_{M}\Gamma_{N},
    \end{equation}
    we can see that an analogous relation to that satisfied in flat space is obtained, given by
    \begin{equation}
        \{\tilde{\Gamma}_{M},\tilde{\Gamma}_{N}\}=2g_{MN}
    \end{equation}

    In particular, for AdS$_5$, the inertial coordinates are $\xi^{A}=\frac{1}{z}x^{A}$, so the vielbein reads
    \begin{equation}
        e^{M}_{N}=\frac{R}{z}\delta^{M}_{N}.
    \end{equation}

    In order to write the Dirac equation in a curved spacetime, we need to replace the partial derivative in (\ref{minkDE}) by a covariant derivative. The covariant derivative of a spinor is given by
    \begin{equation}\label{covariantderivative}
        D_{M}=\partial_{M}+\frac{1}{4}\omega^{AB}_{M}\Sigma_{AB},
    \end{equation}
    where $\Sigma_{AB}=\frac{1}{2}[\Gamma_{A},\Gamma_{B}]$ is the generator for Lorentz transformations of a spinor. The $\omega^{AB}_{M}$ is the spin connection and indicates the behaviour of the vielbein if transported.
    \begin{equation}\label{spinconnection}
        \omega^{AB}_{M}=\frac{1}{2}\bigg(e^{A}_{N}\partial_{M}g^{NP}e^{B}_{P}+e^{A}_{N}g^{QR}e^{B}_{R}\Gamma^{N}_{QM}-(A\leftrightarrow B)\bigg).
    \end{equation}

    The only non-vanishing Christoffel symbols are
    \begin{equation}
        \Gamma_{5}^{55}=A'(z)\quad;\quad\Gamma_{MN}^{5}=-A'(z)\eta_{MN}\quad;\quad\Gamma_{5N}^{M}=A'(z)\delta^{M}_{N}.
    \end{equation}

    So, the non-vanishing spin connection is
    \begin{equation}
        \omega^{5A}_{M}=-\omega^{A5}_{M}=-A'(z)\delta^{A}_{M}.
    \end{equation}

    Substituting these results in (\ref{covariantderivative}), we obtain
    \begin{equation}
        D_{5}=\partial_{5}\quad;\quad D_{\mu}=\partial_{\mu}-\frac{1}{2z}\Gamma_{\mu}\Gamma_{5}.
    \end{equation}

    Finally, in analogy with (\ref{minkDE}), we write the Dirac equation in AdS$_5$
    \begin{equation}\label{adsDE}
        \big(\tilde{\Gamma}^{M}D_{M}-m_5\big)\psi(x,z)=\bigg(\gamma^{\mu}\partial_{\mu}-\frac{2}{z}\gamma_{5}+\gamma_{5}\partial_{z}-\frac{m_5R}{z}\bigg)\psi(x,z)=0.
    \end{equation}

    Now, we make an \textit{ansatz} of plane wave solution
    \begin{equation}
        \psi(x,z)=e^{-ip\cdot x}u(p)\phi(z),
    \end{equation}
    where the Dirac spinors follow  
    \begin{equation}
        \gamma^{\mu}\partial_{\mu}[e^{-ip\cdot x}u(p)]=M[e^{-ip\cdot x}u(p)],
    \end{equation}
    with $M$ the mass of the fermions in the 4-dimensional flat boundary. Let us split the solutions by chirality in left and right $(+/-)$ modes,
    \begin{equation}
        \psi_{\pm}=e^{-ip\cdot x}u_{\pm}(p)\phi_{\pm}(z).
    \end{equation}

    By definition, we have that $\gamma_{5}u_{\pm}=\pm u_{\pm}$ and the following relation holds:
    \begin{equation}
        \gamma^{\mu}\partial_{\mu}(e^{-ip\cdot x}u_{\pm})=M(e^{-ip\cdot x}u_{\mp}).
    \end{equation}

    Back with these definitions and relations in (\ref{adsDE}), we find
    \begin{equation}
        \bigg(\gamma^{\mu}\partial_{\mu}-\frac{m_5 R}{z}\bigg)\bigg(\psi_{+}+\psi_{-}\bigg)+\bigg(\gamma_{5}\partial_{z}-\frac{2}{z}\gamma_{5}\bigg)\bigg(\psi_{+}+\psi_{-}\bigg)=0,
    \end{equation}
    so,
    \begin{equation}
        M\bigg(\phi_{+}u_{-}+\phi_{-}u_{+}\bigg)-\frac{m_{5}R}{z}\bigg(\phi_{+}u_{+}+\phi_{-}u{-}\bigg)+\bigg(\partial_{z}-\frac{2}{z}\bigg)\bigg(\phi_{+}u_{+}-\phi_{-}u_{-}\bigg)=0.
    \end{equation}

    Comparing the terms with chirality left, we find
    \begin{equation}\label{leftDE}
        \bigg(\partial_{z}-\frac{2+m_{5}R}{z}\bigg)\phi_{+}=-M\phi_{-}.
    \end{equation}

    Similarly, the right spinors give us
    \begin{equation}\label{rightDE}
        \bigg(\partial_{z}-\frac{2-m_{5}R}{z}\bigg)\phi_{-}=M\phi_{+}.
    \end{equation}

    The equations (\ref{leftDE}) and (\ref{rightDE}) can be used together to write two equations of second order,
    \begin{equation}\label{SchoeComp}
        \bigg[\partial_{z}^2 -\frac{2}{z}+\frac{5^2-4(m_{5}R\mp 1/2)^2}{4z^2}+M^2\bigg]\phi_{\pm}(z)=0.
    \end{equation}
     Defining $\nu_{\pm}=(m_{5}R\mp1/2)$, and  making the  substitution $\phi_{\pm}(z)=z^{5/2}f_{\pm}(z)$, the equations become
    \begin{equation}
        z^{5/2}\big[z^2\partial_{z}^2+z\partial_{z}+(M^2z^2-\nu_{\pm}^2)\big]f_{\pm}(z)=0.
    \end{equation}
    These equations for $f_{\pm}(z)$ are the Bessel equations whose solutions are linear combinations of the Bessel and Neumann functions of order $\nu_{\pm}$. We are interested in regular solutions at $z=0$, so we take the coefficients related to Neumann solutions as zero for all modes; in this way, the solutions are
    \begin{equation}
        \phi_{\pm}(z)=C_{\pm} z^{5/2}J_{\nu_{\pm}}(z).
    \end{equation}

    For fermions, the connections between the scaling dimension and the mass in the bulk are given by
    \begin{equation}\label{m5delta}
        \Delta= 2 + |m_5 R|,
    \end{equation}
    and baryons with spin $S=1/2$ in four dimensions, the mass dimension is
    \begin{equation}\label{canmassdim1}
        \Delta=\frac{9}{2}+L.
    \end{equation}
Combining Eqs. \eqref{m5delta} and \eqref{canmassdim1}, one finds $|m_5R|=L+5/2$, which substituted in Eq. \eqref{SchoeComp} gives $\nu_{-}=\alpha+1$ and $\nu_{+}=\alpha$, where we  defined $\alpha = 2+L$. With these definitions, we find the full plane wave solution in terms of right and left chirality
    \begin{equation}\label{diracsol}
        \psi_{\alpha}(x,z)=e^{-ip\cdot x}z^{5/2}\big[C_{\alpha}J_{\alpha}( m_{\alpha}z)u_{+}(p)+C_{\alpha+1}J_{\alpha+1}( m_{\alpha+1}z)u_{-}(p)\big],
    \end{equation}
    where the parameters $m_{\alpha}$ and $m_{\alpha+1}$ have dimension of mass, such that the arguments of the Bessel functions are dimensionless. 

    Before we proceed implementing the phenomenological model and comparing to experimental data, let us discuss higher half-integer spin fields.

    \subsection{Fermions with spin $3/2$}

    Fermions with higher spins can be described by objects that mix spinor and tensor indices \cite{Rarita:1941mf}, $\psi_{\beta,\{\mu_1...\mu_{n}\}}$, where $\beta$ is the spinor index; $\mu_i$ are vector indices, and $\{...\}$ means symmetrization of the indices. To describe fermions with spin 3/2, we have the Rarita-Schwinger equation. In the original paper, their expression suggests an alternative form given by
    \begin{equation}
        (\gamma^{[\mu}\gamma^{\nu}\gamma^{\lambda]}\partial_{\nu}+m\gamma^{[\mu}\gamma^{\lambda]})\psi_{\lambda}=0,
    \end{equation}
    where $[...]$ means anti-symmetrization of the indices. This form can be extended to non-Euclidean geometry
    \begin{equation}
    \big(\tilde{\Gamma}^{[M}\tilde{\Gamma}^{N}\tilde{\Gamma}^{L]}\partial_{N}+m\tilde{\Gamma}^{[M}\tilde{\Gamma}^{L]}\big)\psi_{L}=0.
    \end{equation}

    Introducing
    \begin{equation}
        \hat{\psi}_{A}=e^{M}_{A}\Psi_{M},
    \end{equation}
    and choosing $\Psi_{5}=0$ it has been shown \cite{Rarita:1941mf} that $\hat{\psi}_{A}$ satisfies indeed an equation of the form (\ref{adsDE}). It is
    \begin{equation}
        (\gamma^{\nu}\partial_{\nu}-m)\psi_{\{\mu_1...\mu_{n}\}}=0\quad;\quad\gamma^{\mu_1}\psi_{\{\mu_1...\mu_{n}\}}=0.
    \end{equation}

    Indeed, the general solution can be obtained from the Dirac equation with covariant derivative, but all tensor indices referring to the inertial frame, so we introduce 
    \begin{equation}
        \hat{\psi}_{\{A_1...A_n\}}=e^{M_1}_{A_1}...e^{M_n}_{A_n}\psi_{\{A_1...A_n\}},
    \end{equation}
    which satisfy the Dirac equation (\ref{adsDE})
    \begin{equation}
        \big(\tilde{\Gamma}^{M}D_{M}-m\big)\hat{\psi}_{\{A_1...A_n\}}=0.
    \end{equation}

    In anti-de Sitter space, we have $\hat{\psi}_{\{A_1...A_n\}}=z^{n}\psi_{\{A_1...A_n\}}$. In particular, for spin 3/2 we have $\hat{\psi}_{\mu}=z\psi_{\mu}$, substituting this relation in (\ref{adsDE}), we obtain equations similar to the spin 1/2 case. The second order equations become
    \begin{equation}
        \bigg[\partial_{z}^2 -\frac{3}{z}+\frac{7^2-4(m_{5}R\mp 1/2)^2}{4z^2}+M^2\bigg]\phi_{\pm}(z)=0, 
    \end{equation}
    and the solutions can be written as 
    \begin{equation}
            z^{7/2}\big[z^2\partial_{z}^2+z\partial_{z}+(M^2z^2-\nu_{\pm}^2)\big]f_{\pm}(z)=0.
    \end{equation}

    Here, since we are considering a spin $S=3/2$ particle as a sum of a spin $S=1$ and another $S=1/2$, we will interpret such fields as a vector field and a spin-$1/2$ field, which leads us to the dimension
       \begin{equation}\label{canmassdim2}
        \Delta=\frac{11}{2}+L,
    \end{equation}
    whose solutions will be
    \begin{equation}\label{raritasol}
        \psi_{\alpha}^{\mu}(x,z)=\epsilon^{\mu}e^{-ip\cdot x}z^{7/2}\big[C_{\alpha+1}J_{\alpha+1}( m_{\alpha+1}z)u_{+}(p)+C_{\alpha+2}J_{\alpha+2}( m_{\alpha+2}z)u_{-}(p)\big], 
    \end{equation}
where $\alpha=2+L$, as in the case of spin 1/2 fermions. Note that here the positive parity index for the Bessel functions is $\nu_+=\alpha+1$, while for the negative parity it is $\nu_-=\alpha+2$. With these solutions, we can construct the HW mass tower.


   \FloatBarrier

\section{A Hard wall model and Anomalous versions}

  In this section, we discuss the holographic Hard Wall model (HW), based on the idea previously developed in
  \cite{Boschi-Filho:2002xih, Boschi-Filho:2002wdj, deTeramond:2005su, Erlich:2005qh, Boschi-Filho:2005xct} and propose an improved version of the Anomalous Hard Wall model (AHW) for baryons, previously developed for glueballs and light mesons \cite{Costa-Silva:2023vuu,Costa-Silva:2024dlq}. Unlike the original HW model for baryons \cite{deTeramond:2005su}, this version differs since we assign distinct mass dimensions to spin-$1/2$ and spin-$3/2$ objects.

  The main idea of the hard wall model is to introduce a hard cut off in the AdS space breaking conformal invariance. This was proposed by Polchinski and Strassler to study the hard scattering of glueballs \cite{Polchinski:2001tt} and was adopted to calculate scalar glueball masses in Refs. \cite{Boschi-Filho:2002xih, Boschi-Filho:2002wdj}. In Poincaré coordinates, Eq. \eqref{metric}, one can set  
  \begin{equation}
         \frac{1}{z_{\rm max}}=\Lambda_{\rm HW},
    \end{equation}
where in this model, the holographic coordinate $z$ is now defined in the range $[0,z_{\rm max}]$. This sets a mass/energy scale for the hard wall model. Then, a boundary condition is imposed in $z=z_{\rm max}$ descretizing the field wave functions implying discrete mass spectra, and the mass parameters $m_{\alpha}$ become related to the $k$-th zeros $\chi_{\alpha, k}$ of a given Bessel function $J_\alpha (m_{\alpha, k} z)$.  
This model was later extended to mesons and baryons \cite{deTeramond:2005su, Erlich:2005qh} and higher spin glueballs \cite{Boschi-Filho:2005xct}. 
Since in this work we are not considering radial excitations, the masses of the baryons will all have fixed $k=1$, which we are going to omit from now on. 



  \subsection{Hard Wall Model}
  
  As discussed in Section \ref{Review:fermions}, fermionic solutions with spin 1/2 and 3/2, Eqs. (\ref{diracsol}) and (\ref{raritasol}), are described by two modes, each with positive and negative parities. In order to obtain the masses associated with each of the modes, we now impose boundary conditions at $z=z_{\rm max}$ on these solutions; in particular, we impose Dirichlet boundary conditions. Clearly, we cannot simultaneously impose such conditions on both modes, so each mode has its own spectrum. As in \cite{deTeramond:2005su}, the right-handed mode will be associated with the positive-parity spectrum, while the left-handed mode will correspond to the negative-parity spectrum. The parity $P$ of the states is connected to their  orbital angular momentum $L$. Particles with even (odd) $L$ have positive (negative) parity. 
  
  Thus, for the spin-$1/2$ we have from Eq. (\ref{diracsol}) and setting the positive parity so that $\nu=\nu_+=\alpha=2+L$ we obtain,
\begin{equation}
        M_{1/2+}(L)=\chi_{2+L}\Lambda_{\rm HW};
        \qquad{\rm (nucleons).}
        \label{M_{+;1/2}HW}
    \end{equation}
 This equation applies for nucleons since these particles have spin 1/2 states with positive parity, which is not the case for deltas, as can be seen from PDG data compiled in Tables \ref{pdgN} and \ref{pdgDelta}. 

For spin-$1/2$ and negative parity, we have from Eq. (\ref{diracsol}) and $\nu=\nu_-=\alpha+1=3+L$, so that  
    \begin{equation}
        M_{1/2-}(L)=\chi_{3+L}\Lambda_{\rm HW};
        \qquad{\rm (nucleons\; and\; deltas).}
        \label{M_{-;1/2}HW}
    \end{equation}
Note that this equation applies equally well for nucleons and deltas since both particles have states with spin 1/2 and negative parity, as can be checked in Tables \ref{pdgN} and \ref{pdgDelta}. 

For spin-$3/2$ the solution comes from the Rarita-Schwinger equation, Eq. (\ref{raritasol}). Then, choosing the positive parity solution $u_+$, we have $\nu=\nu_+=\alpha+1=3+L$, so that 
  \begin{equation}
        M_{3/2+}(L)=\chi_{3+L}\Lambda_{\rm HW};
        \qquad{\rm (nucleons\; and\; deltas),}
        \label{M_{+;3/2}HW}
    \end{equation}
which applies to nucleons and deltas, according to PDG data, Tables \ref{pdgN} and \ref{pdgDelta}.  

For spin-3/2 and negative parity, one still ressources to Eq. (\ref{raritasol}) with $\nu=\nu_-=\alpha+2=4+L$, then 
  \begin{equation}
        M_{3/2-}(L)=\chi_{4+L}\Lambda_{\rm HW};\qquad{\rm (nucleons),}
        \label{M_{-;3/2}HW}
    \end{equation}
which applies only to nucleons, once deltas do not have spin 3/2 with negative parity. 

  Note also that the masses for negative parity with spin 1/2, Eq. \eqref{M_{-;1/2}HW} and positive parity with spin 3/2, Eq. \eqref{M_{+;3/2}HW}, coincide, both for nucleons and deltas, so these states and the corresponding Regge trajectories will be degenerate. 
  
  The mass scale $\Lambda_{\rm HW}$ is chosen to obtain the best fit of the masses tabulated by PDG. Then, we look for the value that minimizes the function $Q$, defined in equation (\ref{qfunction}). This is done separately for nucleons and deltas implying two values of $\Lambda_{\rm HW}$, one for each family. We will solve this problem numerically in Section \ref{numericalresults}. The results are shown in Tables  \ref{tabdeltas} and \ref{tabnucleons}. The corresponding Regge trajectories obtained from this version of the hard wall model are shown in Fig. \ref{figHW}.


\FloatBarrier
\subsection{Anomalous Hard Wall Models}

Now, we proceed to discuss the anomalous hard wall model for baryons. It is a well known fact that in an interacting field theory, the anomalous dimension plays an important role due to renormalization. Since baryons are made of interacting three valence quarks, it is reasonable that the mass dimension of an operator that corresponds to a baryon state is not simply the sum over canonical dimensions (free theory) of such quarks but is corrected by the anomalous dimension. In order to treat the dimension of such operators we will write the total dimension, $\Delta$ as a sum of the canonical part (no interacting theory) and the anomalous one. So, we consider that the total dimension of a hadronic operator $(\Delta_{\rm total})$ is the sum of its canonical dimension $(\Delta_{\rm can.})$ plus its anomalous dimensions $(\Delta_{\rm anom.})$: 
\begin{equation}\label{totaldim}
    \Delta_{\rm total} = \Delta_{\rm can.} + \Delta_{\rm anom.}. 
\end{equation}

The canonical dimension $\Delta_{\rm can.}$ was discussed in Sect. \ref{Review:fermions} for spins 1/2 and 3/2, in Eqs. \eqref{canmassdim1} and \eqref{canmassdim2}, respectively, 
\begin{eqnarray}\label{deltacan}
\Delta_{\rm can.}= \left\{
\begin{array}{lccccc}
\displaystyle\frac{9}{2}+L; &&&&& S=1/2, \cr  \\ 
\displaystyle\frac{11}{2}+L; &&&&&  S = 3/2.
\end{array}
\right. 
\end{eqnarray}
It is important to note that both nucleons and deltas can have spin-1/2 or spin-3/2. Therefore, whether the canonical dimension will be given by the first or second of the equations (\ref{deltacan}) will depend only on its spin and not on the family of the particle. 

From a semiclassical gauge/gravity duality \cite{Gubser:2002tv}, the anomalous dimension of an operator at the boundary increases logarithmically with its total angular momentum $J$
\begin{equation}\label{anomalousdimension}
    \Delta_{\rm anomalous}=\frac{\sqrt{\lambda}}{\pi}\ln{\bigg(\frac{J}{\sqrt{\lambda}}\bigg)}+\mathcal{O}(J^0),
\end{equation}
where $\lambda=g_{\rm YM}^2N$ is the 't Hooft coupling for an $SU(N)$ Yang-Mills theory. The above relation is valid when $\lambda \gg 1$.

Inspired by this result, we will propose anomalous corrections to the hard wall model. First, we suppose that the anomalous dimension depends only on the orbital angular momentum excitation $L$, and is given by
\begin{equation}\label{deltaanom1}
\Delta_{\rm anom.1} = a\ln{(L+1)}+b, 
\end{equation}
where $a$ and $b$ are free parameters to be adjusted to fit PDG data. Note that we have chosen the argument of the logarithm as $L+1$ in order to avoid a singularity in the 
the ground state $L=0$. 

In our second proposal, the anomalous dimension depends on $L$, the orbital angular momentum excitation, and $S$, the intrinsic spin. To make (for $S=1/2$) the anomalous dimension related to the first case, we write
\begin{equation}\label{deltaanom2}
    \Delta_{\rm anom.2} = a\ln{(L+S+1/2)}+b, 
\end{equation}
where $a$ and $b$ are again free parameters that will be fixed later. 

So, one can go back to the discussion of Section \ref{Review:fermions} and substitute the canonical dimensions for spins 1/2 and 3/2, Eqs. \eqref{canmassdim1} and \eqref{canmassdim2}, respectively, by the total dimension given by Eq. \eqref{totaldim}, with the anomalous dimensions given by \eqref{deltaanom1} or \eqref{deltaanom2}, which we call, respectively, AHW1 and AHW2. This implies that we replace $\alpha=2+L$ by $\alpha=2+L+\Delta_{\rm anom.1}$ or by $\alpha=2+L+\Delta_{\rm anom.2}$.

The energy scale parameter, previously denoted by $\Lambda_{\rm HW}$, which here will be denoted as $\Lambda_{\rm AHW}$, is obtained as before: minimizing the function in (\ref{qfunction}); for simplicity, we will make this parameter the same for the two versions of the anomalous dimension analyzed. 

The masses of the positive-parity spin states $1/2$ from Eq. (\ref{diracsol}) and $\nu=\nu_+=\alpha=2+L+\Delta_{\rm anom.}$ we obtain for nucleons
\begin{equation}
    M_{1/2+}(L)=\chi_{2+L+\Delta_{\rm anom.}}\Lambda_{\rm AHW};
    \qquad(\rm nucleons), 
    \label{M_{+;1/2}AHW}
\end{equation}
and for spin 1/2 and negative parity with $\nu=\nu_-=\alpha+1=3+L+\Delta_{\rm anom.}$, one has for nucleons and deltas
\begin{equation}
    M_{1/2-}(L)=\chi_{3+L+\Delta_{\rm anom.}}\Lambda_{\rm AHW};\qquad(\rm nucleons\; and \; deltas).
    \label{M_{-;1/2}AHW}
\end{equation}
For spin-$3/2$, with solution given by Eq. (\ref{raritasol}), and positive parity, we have $\nu=\nu_+=\alpha+1=3+L+\Delta_{\rm anom.}$ for nucleons and deltas
\begin{equation}
    M_{3/2+}(L)=\chi_{3+L+\Delta_{\rm anom.}}\Lambda_{\rm AHW};\qquad(\rm nucleons\; and \; deltas), \label{M_{+;3/2}AHW}
\end{equation}
and for spin 3/2 and negative parity, $\nu=\nu_-=\alpha+2=4+L+\Delta_{\rm anom.}$ for nucleons
\begin{equation}
    M_{3/2-}(L)=\chi_{4+L+\Delta_{\rm anom.}}\Lambda_{\rm AHW}; \qquad(\rm nucleons).\label{M_{-;3/2}AHW}
\end{equation}

Note that the masses for negative parity with spin 1/2, Eq. \eqref{M_{-;1/2}AHW} and positive parity with spin 3/2, Eq. \eqref{M_{+;3/2}AHW}, do coincide for  $\Delta_{\rm anom.1}$, Eq. \eqref{deltaanom1}, as in the case of the usual hard wall model, for nucleons and deltas, but not for $\Delta_{\rm anom.2}$, Eq. \eqref{deltaanom2}, as can be seen in Figs. \ref{figAHW1} and \ref{figAHW2}.

The parameters $\Lambda_{\rm AHW}$ in the above equations and $a$, and $b$ in Eqs. \eqref{deltaanom1} and \eqref{deltaanom2}, are obtained by minimizing the function $Q$, Eq. \eqref{qfunction}, which we solve numerically in the next section, showing the results of both possibilities for the proposed anomalous dimensions.  The corresponding masses are shown in Tables \ref{tabdeltas} and \ref{tabnucleons}, and the Regge trajectories in Figs. \ref{figAHW1} and \ref{figAHW2}.


\subsection{Numerical Results for the HW and AHW models}\label{numericalresults}

\begin{figure}[ht!]
\centering
\includegraphics[width=12cm]{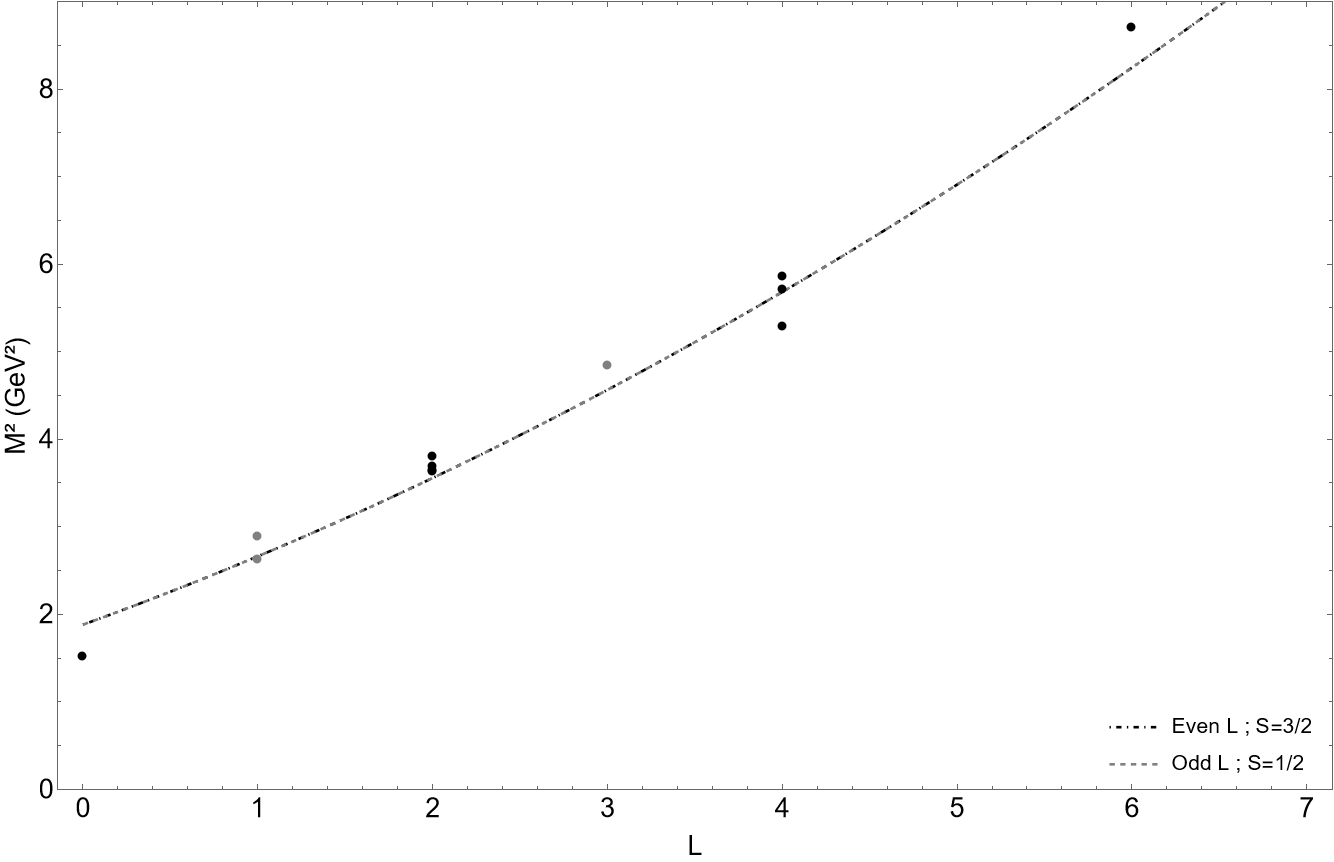}\quad 
\includegraphics[width=12cm]{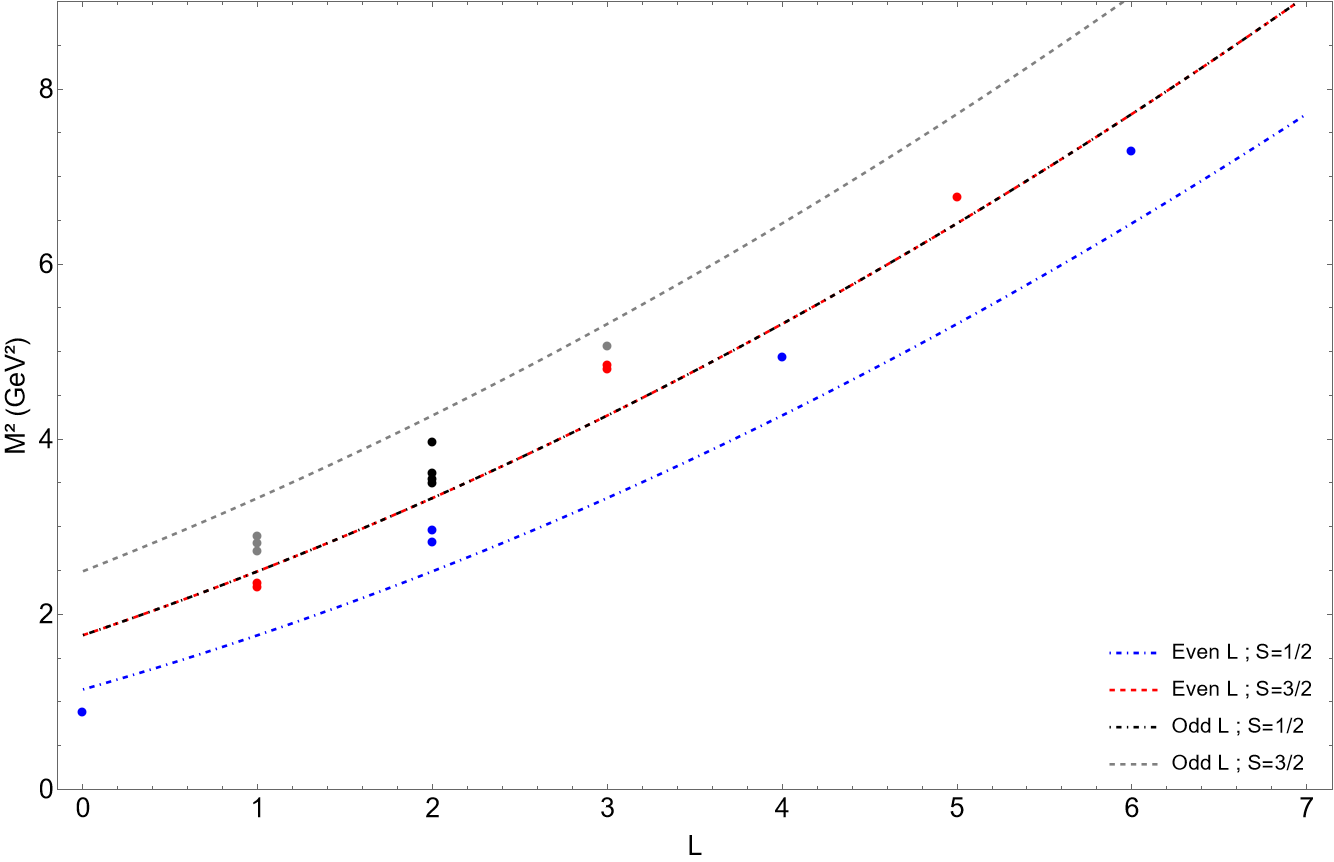}
\caption{Regge trajectories ($M^2\times L$) obtained from the HW model from equations (\ref{M_{+;1/2}HW})-(\ref{M_{-;3/2}HW}), respectively with colors  blue (dot-dashed), black(dot-dashed), red (dashed), and grey (dashed), for delta resonances (upper panel) and nucleon resonances (lower panel), with masses $M$ expressed in GeV. As explained in the text, the delta resonances only present states with spin 1/2 with negative parity and spin 3/2 with positive parity, which in the HW are degenerate. Note that the trajectories corresponding to Eqs. \eqref{M_{-;1/2}HW} (black/dot-dashed) and \eqref{M_{+;3/2}HW} (red/dashed) coincide in both panels. We also present the corresponding experimental values (dots) obtained from the PDG. Each point of a given color corresponds to the trajectory of the same color.}
\label{figHW}
\end{figure}

\begin{figure}[ht!]
\centering
\includegraphics[width=12cm]{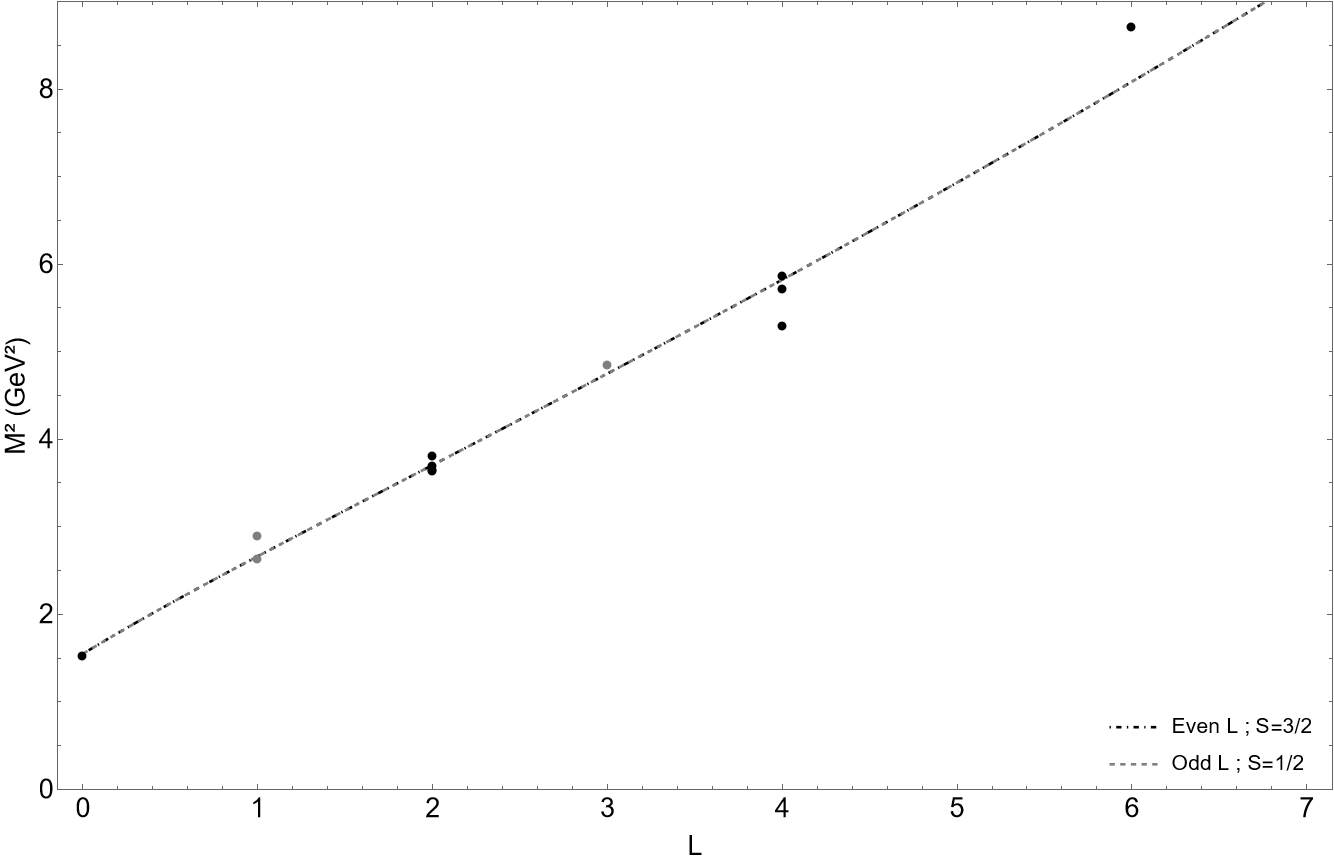}\quad
\includegraphics[width=12cm]{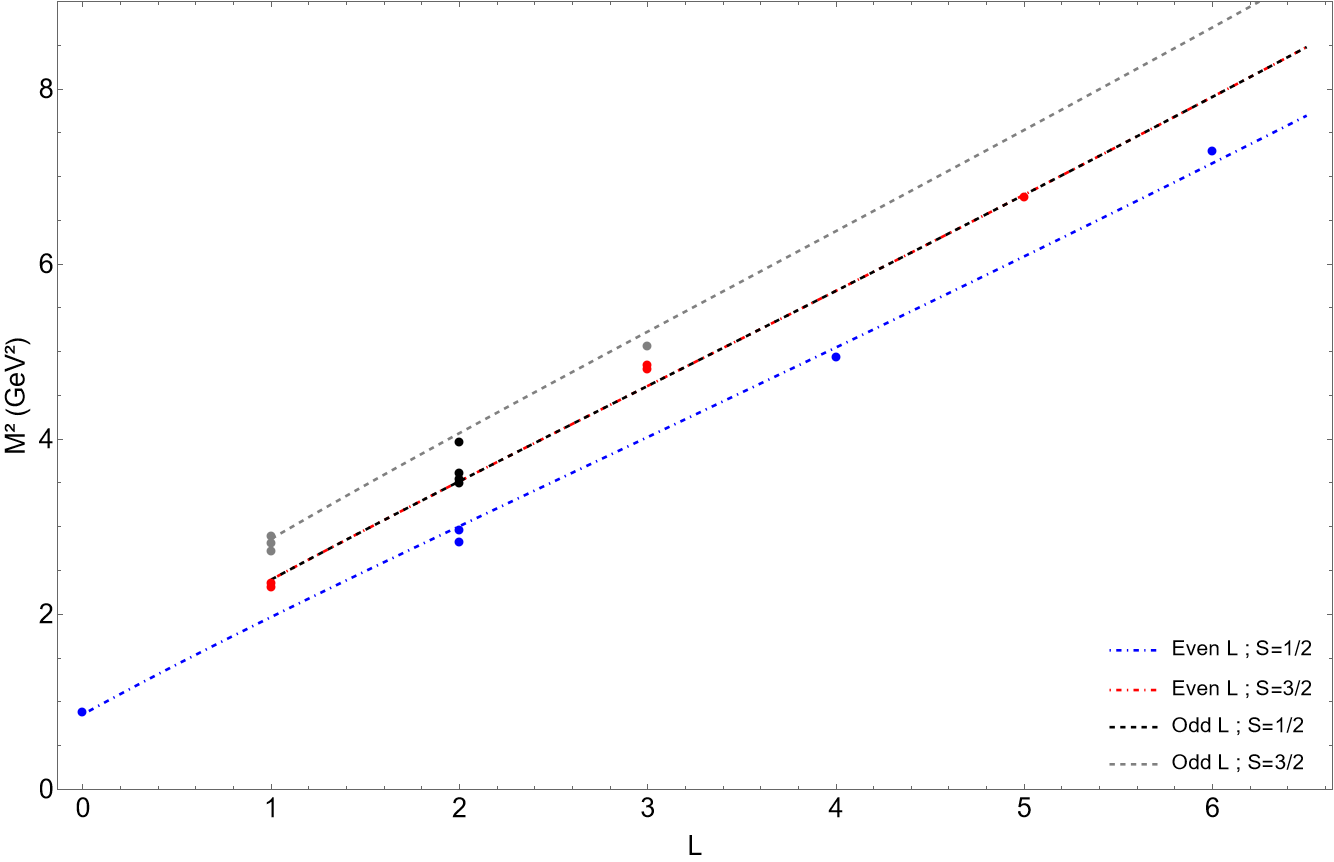}
\caption{Regge trajectories ($M^2\times L$) obtained from the AHW1 model, with anomalous dimension, Eq.  (\ref{deltaanom1}), from equations (\ref{M_{+;1/2}AHW})-(\ref{M_{-;3/2}AHW}), respectively with colors  blue (dot-dashed), black(dot-dashed), red (dashed), and grey (dashed), for delta resonances (upper panel) and nucleon resonances (lower panel), with masses $M$ expressed in GeV. As explained in the text, the delta resonances only present states with spin 1/2 with negative parity and spin 3/2 with positive parity, which in the AHW1 are degenerate. Note that the trajectories corresponding to Eqs. \eqref{M_{-;1/2}AHW} (black/dot-dashed) and \eqref{M_{+;3/2}AHW} (red/dashed) coincide in both panels. We also present the corresponding experimental values (dots) obtained from the PDG. Each point of a given color corresponds to the trajectory of the same color.}
\label{figAHW1}
\end{figure}

    In this section, we present the numerical results for the Hard Wall and Anomalous Hard Wall models proposed in  previous sections  and compare them with PDG data. In order to obtain the mass values from these models given a state with orbital angular momentum $L$ and spin $S$, we need to use the appropriate equation from (\ref{M_{+;1/2}HW})-(\ref{M_{-;3/2}HW}), in the HW case, or from (\ref{M_{+;1/2}AHW})-(\ref{M_{-;3/2}AHW}) for the AHW. As mentioned above, the parameters $a$, $b$ and the energy scale $\Lambda$ are obtained by minimizing the function $Q$ in equation (\ref{qfunction}). Since this function depends on the obtained masses, it is a recursive problem that we solve numerically and present the values of the parameters, as well as the masses for each state and each model.
    
    In Table \ref{tabmodelinfo}, we present the numerical values for $\Lambda$, the energy scale, the values of the parameters $a$ and $b$, and the values of the function $Q$ divided by the number $n$ of states available in PDG, for each model and particle family (nucleons and deltas). Note that the values of $\Lambda$ found for the AHW$_1$ for deltas and nucleons were also used in the AHW$_2$, for these families, for simpicity. In this table, one can see that the total deviation $Q$ is smaller for the AHW$_1$ model compared to the original HW and AHW$_2$ models.

\vskip 0.5cm 
\begin{table}[ht]
\begin{tabular}{|c|c|c|c|c|c|}
\hline
\hline 
Family  & Model&$\Lambda$ (MeV) & $a$& $b$ &  $Q/n $ \%
\\ \hline
Deltas & HW & $215$ &- & - &   (13.9/12)\% = 1.16\%\\
\hline
 Nucleons & HW &  208 & - & - &  (28.6/18)\% = 1.59\%\\
 \hline\hline
Deltas  & AHW$_1$ & 151 & \; 1.74 \; & \; 1.55 \; &   (7.75/12)\% = 0.646\%\\
\hline
 Nucleons &AHW$_1$& 127 & 3.20 & 1.75 &  (8.61/18)\% = 0.478\%\\
 \hline \hline
Deltas  & AHW$_2$ & 151 & 1.90  &  0.89  &   (17.1/12)\% = 1.43\%\\
\hline
 Nucleons & AHW$_2$ & 127 & 2.82 & 1.69 &  (19.2/18)\% = 1.07\%\\
 \hline\hline
\end{tabular}
\caption{\label{tabmodelinfo} Values of the relevant parameters of the models HW, AHW1, and AHW2 obtained by the minimization of $Q$ function, Eq. \eqref{qfunction}. Notice that the HW model does not present $a$ and $b$ parameters. In the last column we present the value of the $Q$ function over the $n$ states used to construct such function.}
\end{table}

    In Tables \ref{tabdeltas} and \ref{tabnucleons}, we present our results for the masses of the delta and nucleon resonances, respectively, for the HW, AHW$_1$, and AHW$_2$ models. The first three columns of both tables show the orbital angular momentum excitation $L$, the parity $P$, and the spin $S$, of each state. The fourth column presents the corresponding masses of these states given by PDG. These Tables also present our results for the masses of these states obtained from the HW model, the AHW$_1$ and AHW$_2$, as well as the percentage deviation of each of these masses compared to PDG. As anticipated in Table \ref{tabmodelinfo}, one can see in Tables \ref{tabdeltas} and \ref{tabnucleons} that the masses obtained from the AHW$_1$ for the delta and nucleon resonances have smaller deviations with respect to PDG data than the ones from the original HW and AHW$_2$ models.

\vskip 0.5cm
\begin{table}[ht]
\begin{tabular}{|c|c|c|c|c|c|c|c|c|c|}

\hline\hline
\;$L$\; & \;$P$\; & \;$S$\; &\;PDG\; & \;HW\; & \;$\delta_{\rm HW}$\;  & \;AHW$_1$\; &$\delta_{\rm{AHW}_1}$  & \;AHW$_2$\;  &$\delta_{\rm{AHW}_2}$ \\\hline
0 & $+$ & $\tfrac{3}{2}$ & 1232 & 1372 & 11.3 &1244& 1.01 & 1361 & 10.5\\
1 & $-$ & $\tfrac{1}{2}$ & 1620 & 1631 & 0.709 & 1632 & 0.731 & 1536 & 5.16 \\
1 & $-$ & $\tfrac{1}{2}$ & 1700 & 1631 & 4.03 & 1632 & 4.01 & 1536 & 9.62 \\
2 & + & $\tfrac{3}{2}$ & 1910 & 1886 & 1.26 & 1925 & 0.778 & 1935 & 1.33\\
2 & + & $\tfrac{3}{2}$ & 1920 & 1886 & 1.78 & 1925 & 0.253 & 1935 &0.807\\
2 & + & $\tfrac{3}{2}$ & 1905 & 1886 & 1.00 & 1925 & 1.04 & 1935 & 1.60\\
2 & + & $\tfrac{3}{2}$ & 1950 & 1886 & 3.29 & 1925 & 1.29 & 1935 & 0.744\\
3 & $-$ & $\tfrac{1}{2}$ & 2200 & 2136 & 2.90 & 2179 & 0.936 & 2105 & 4.30\\
4 & + & $\tfrac{3}{2}$ & 2390 & 2384 & 0.269 & 2413 & 0.955 & 2403 & 0.561\\
4 & + & $\tfrac{3}{2}$ & 2300 & 2384 & 3.63 & 2413 & 4.90 & 2403 & 4.50\\
4 & + & $\tfrac{3}{2}$ & 2420 & 2384 & 1.51 & 2413 & 0.297 & 2403 & 0.685 \\
6 & + & $\tfrac{3}{2}$ & 2950 & 2871 & 2.67 & 2843 & 3.62 & 2828 & 4.15\\
\hline\hline
\end{tabular}
\caption{\label{tabdeltas}  Delta resonance  states characterized by their orbital angular momentum $L$, parity $P$, and spin $S$, with their corresponding masses (in MeV) from PDG, and our results from the HW, Eqs.  (\ref{M_{+;1/2}HW})-(\ref{M_{-;3/2}HW}), and  AHW$_1$ and AHW$_2$ models, Eqs. (\ref{M_{+;1/2}AHW})-(\ref{M_{-;3/2}AHW}) with $\Delta_{\rm anom.}$ given by Eqs (\ref{deltaanom1}) and (\ref{deltaanom2}), respectively. For comparison, we also present the percentage deviation between these masses and the ones from PDG.}
\end{table}

\begin{table}[ht]
\begin{tabular}{|c|c|c|c|c|c|c|c|c|c|}
\hline\hline
\;$L$\; &\;$P$\; & \;$S$\; & \; PDG \; & \; HW \; & \;$\delta_{\rm{HW}}$\; & \;AHW$_1$\; & $\delta_{\rm{AHW}_1}$   & \;AHW$_2$\; & $\delta_{\rm{AHW}_2}$ \\ \hline
0 & + & $\tfrac{1}{2}$ & 939 & 1068 & 13.8 & 926 & 1.42 & 916 & 2.39 \\
1 & $-$ & $\tfrac{1}{2}$ & 1520 & 1578 & 3.84 & 1548 & 1.84 & 1501 & 1.23\\
1 & $-$ & $\tfrac{1}{2}$ & 1535 & 1578 & 2.83 & 1548 &0.846 & 1501 & 2.19\\
1 & $-$ & $\tfrac{3}{2}$ & 1650 & 1824 & 10.6 & 1691 & 2.51 & 1808 & 9.59\\
1 & $-$ & $\tfrac{3}{2}$ & 1700 & 1824 & 7.32 & 1691 & 0.504 & 1808 & 6.37\\
1 & $-$ & $\tfrac{3}{2}$ & 1675 & 1824 & 8.92 & 1691 & 0.981 & 1808 & 7.96\\
2 & + & $\tfrac{1}{2}$ & 1720 & 1578 & 8.23 & 1734 & 0.808 & 1665 & 3.16\\
2 & + & $\tfrac{1}{2}$ & 1680 & 1578 & 6.05 & 1734 & 3.21 & 1665 & 0.851\\ 
2 & + & $\tfrac{3}{2}$ & 1880 & 1824 & 2.95 &1876 & 0.210 & 1923 & 2.30\\
2 & + & $\tfrac{3}{2}$ & 1900 & 1824 & 3.98 & 1876 & 1.26 & 1923& 1.23\\
2 & + & $\tfrac{3}{2}$ & 1870 & 1824 & 2.43 & 1876 & 0.324 & 1923 & 2.85\\
2 & + & $\tfrac{3}{2}$ & 1990 & 1824 & 8.32 & 1876 & 5.72 & 1923 & 3.35\\
3 & $-$ & $\tfrac{1}{2}$ & 2190 & 2067 & 5.63 & 2147 & 1.97 & 2064 & 5.73\\ 
3 & $-$ & $\tfrac{1}{2}$ & 2220 & 2067 & 6.90 & 2147 & 3.29 & 2064 & 7.01\\
3 & $-$ & $\tfrac{3}{2}$ & 2250 & 2306 & 2.49 & 2247 & 1.64 & 2292& 1.89\\
4 & + & $\tfrac{1}{2}$ & 2220 & 2067 & 6.90 & 2606 & 1.21 & 2152&3.03\\
5 & $-$ & $\tfrac{1}{2}$ & 2600 & 2543 & 2.20 &2606 & 0.244 & 2503&3.71 \\
6 & + & $\tfrac{1}{2}$ & 2700 & 2543 & 5.82 &2675 & 0.939 & 2564&5.04 \\
 \hline \hline
\end{tabular}
\caption{\label{tabnucleons} 
Nucleon resonance  states characterized by their orbital angular momentum $L$, parity $P$, and spin $S$, with their corresponding masses (in MeV) from PDG, and our results from the HW, Eqs.  (\ref{M_{+;1/2}HW})-(\ref{M_{-;3/2}HW}), and  AHW$_1$ and AHW$_2$ models, Eqs. (\ref{M_{+;1/2}AHW})-(\ref{M_{-;3/2}AHW}) with $\Delta_{\rm anom.}$ given by Eqs (\ref{deltaanom1}) and (\ref{deltaanom2}), respectively. For comparison, we also present the percentage deviation between these masses and the ones from PDG.}
\end{table}

It is important to note that in equations (\ref{M_{+;1/2}HW})-(\ref{M_{-;3/2}HW}) for the HW model, there are degeneracies in masses for different $L$, for example, for positive parity, with spin $1/2$ and $L=2$ which presents the same mass as the negative parity, with spin $1/2$ and $L=1$. There are other degeneracies and they appear in the spectrum presented in Tables \ref{tabdeltas} and \ref{tabnucleons}. Even in the AHW$_1$ case, with equations (\ref{M_{+;1/2}AHW})-(\ref{M_{-;3/2}AHW}), the same degeneracies arise. It is a feature of the model, but also expected, since for a given mass, there are states with different angular momentum: those who fall in the leading Regge trajectory and others in the daughters trajectories. For example, in Figures 2 and 8 of reference \cite{Klempt:2012fy}, we can also see this happening for delta and nucleon resonances, respectively.

 Given the masses and, for example, the orbital angular momentum $L$, we can construct the Regge trajectory $M^2\times L$. It is important to note that only states with positive parity and $S=3/2$ or negative parity and $S=1/2$ are known for delta resonances, so we will only construct Regge trajectories in delta cases for the observed states that are degenerated even in the original HW or in the AHW$_1$ and AHW$_2$ models. The "missing states" are discussed, for example, in \cite{Klempt:2009pi}.

\begin{figure}[ht!]
\vskip0.5cm 
\centering
\includegraphics[width=12cm]{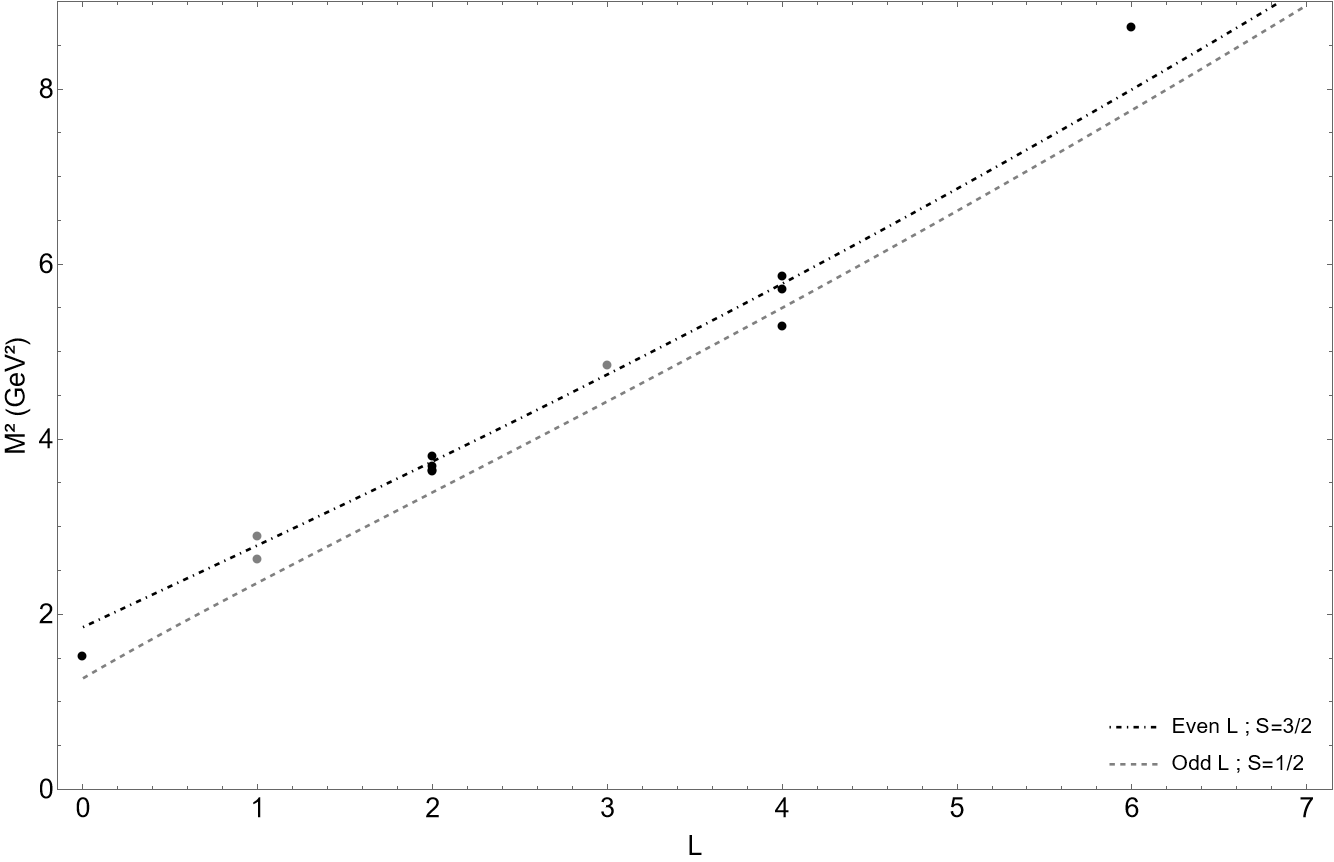}\quad
\includegraphics[width=12cm]{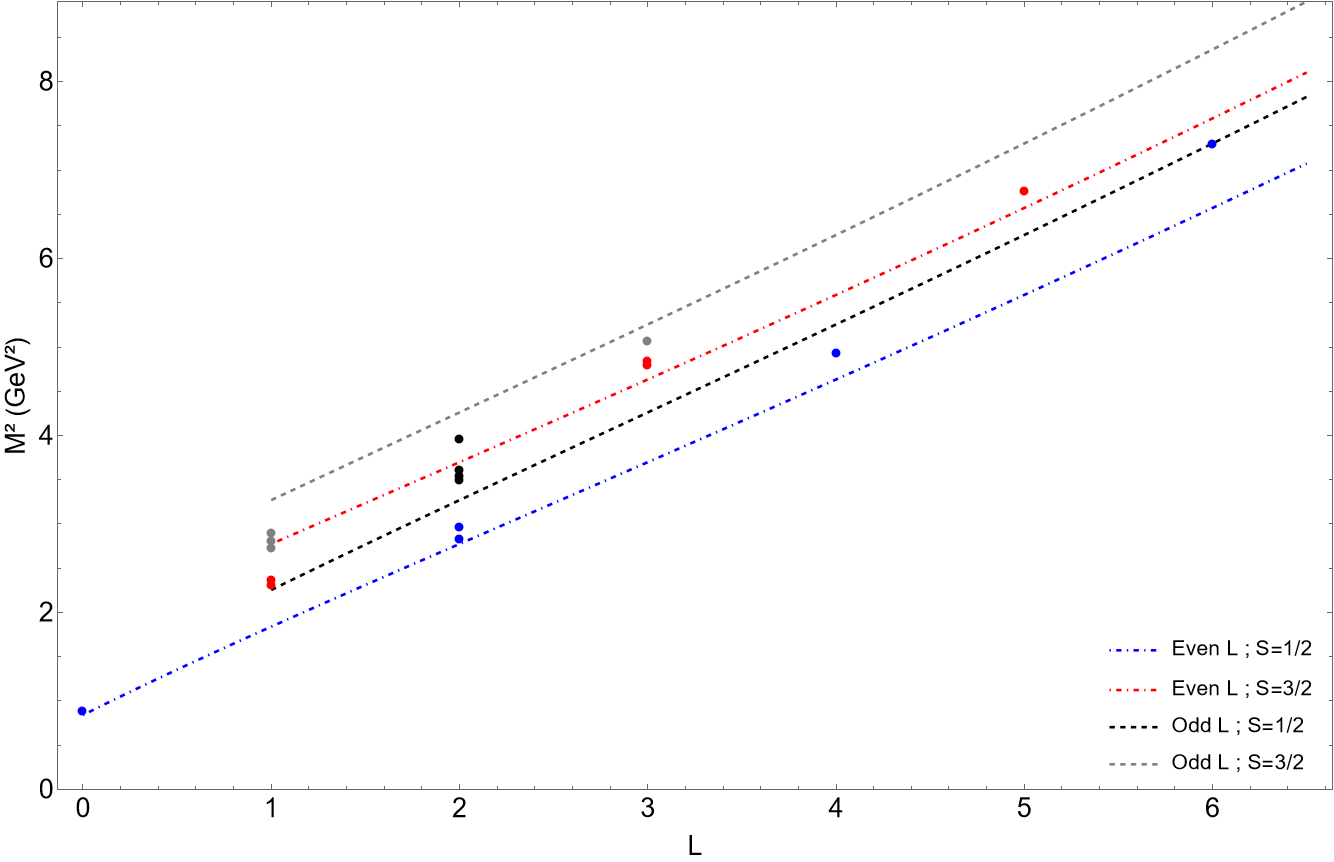}
\caption{Regge trajectories ($M^2\times L$) obtained from the AHW2 model, with anomalous dimension, Eq.  (\ref{deltaanom2}), from equations (\ref{M_{+;1/2}AHW})-(\ref{M_{-;3/2}AHW}), respectively with colors  blue (dot-dashed), black(dot-dashed), red (dashed), and grey (dashed), for delta resonances (upper panel) and nucleon resonances (lower panel), with masses $M$ expressed in GeV. As explained in the text, the delta resonances only present states with spin 1/2 with negative parity and spin 3/2 with positive parity. In both panels, we also present the corresponding experimental values (dots) obtained from the PDG. Each point of a given color corresponds to the trajectory of the same color.}
\label{figAHW2}
\end{figure}

    In Figs. \ref{figHW}, \ref{figAHW1}, and \ref{figAHW2}, we present the Regge trajectories (dashed and dot-dashed lines) obtained from the original HW, AHW$_1$ and AHW$_2$ models, respectively. In the upper panels, we show delta resonances and in the lower panels, nucleon resonances. In all panels, we also present the states obtained from PDG, which are represented by dots. Ideally, each dot of a given color would fall in the mass trajectory of the same color.

    As a well-known result, the trajectories in the original HW model, Fig. \ref{figHW}, do not present the expected linear growth of $M^2$ with $L$. On the other hand, in the trajectories obtained from the AHW$_1$ and AHW$_2$ models, presented in Figs. \ref{figAHW1}, and \ref{figAHW2}, we can see  improvements of two kinds with respect to the HW model: the first is that the masses of PDG lie much closer to the trajectories in these cases. The second is that these  trajectories look much more linear. 
    
    Furthermore, in the case of Fig. \ref{figHW}, representing our results for the HW model, we can see from equations (\ref{M_{-;1/2}HW}) and (\ref{M_{+;3/2}HW}), the mass trajectory for spin $S=1/2$ and negative parity is degenerate with $S=3/2$ and positive parity, both for nucleons and deltas. A similar situation occurs for the AHW$_1$ model with trajectories shown in Fig. \ref{figAHW1}. In this case, the degeneracy is related to Eqs. \eqref{M_{-;1/2}AHW} and \eqref{M_{+;3/2}AHW}. A different behavior happens for the AHW$_2$, with trajectories represented in Fig. \ref{figAHW2}. In this case, there is no degeneracy between the trajectories, since the anomalous dimension given by Eq. (\ref{deltaanom2}),  depends on the angular momentum $L$ and the spin $S$ of the states, and then these degeneracies are broken. So, there are no  crossings between the trajectories in this case.

\section{Linear Anomalous Hard Wall}

An important fact about the Hard Wall model, and one that motivated the development of other holographic models, is that it presents non-linear Regge trajectories. As observed in the previous section, the anomalous models presented much more linear trajectories compared to the original HW model. Despite this, if we extrapolate the plots to observe higher values of angular momentum, we would observe a tendency towards non-linearity.

Based on previously developed ideas, \cite{Costa-Silva:2023vuu,Costa-Silva:2024dlq}, we sought to determine if it is possible to obtain asymptotically linear Regge trajectories for baryons through direct modifications to the total mass dimension. As observed in previous works, a dimension of the type 
\begin{equation}\label{ALHW}
\Delta_{\rm Lin.}= aL^{c}+b,
\end{equation}
leads to asymptotically linear trajectories for even-spin glueballs and light mesons, with $c \approx 0.50$. Here, for baryons, the Regge trajectories obtained are approximately linear for $c\approx 0.75$. A possible explanation is that even glueballs and mesons have two valence particles and in baryons three. So, it could be that one has to sum 0.25 for each valence parton inside the hadron to get a linear Regge trajectory in this model. 

 Taking into account the differences of the previous models with respect to the spin of the resonances, in this model, we propose that the dimension of the baryonic operator is given by
\begin{eqnarray}\label{deltalin}
\Delta = \left\{
\begin{array}{lccccc}
\displaystyle\frac{9}{2}+\Delta_{\rm Lin.}; &&&&& S=1/2, \cr  \\ 
\displaystyle\frac{11}{2}+\Delta_{\rm Lin.}; &&&&&  S = 3/2,
\end{array}
\right. 
\end{eqnarray}
where $S$ is the spin of the baryon. 

 As before, we can write the mass spectrum from the mass dimension of the operators, depending on the parity (whether it has even or odd angular momentum) and the spin. The masses of the positive-parity spin states $1/2$ are then
\begin{equation}
    M_{1/2+}(L)=\chi_{2+\Delta_{\rm Lin.}}\Lambda_{\rm ALHW}; \qquad {\rm (nucleons)},\label{M_{+;1/2}ALHW}
\end{equation}
and for negative parity,
\begin{equation}
    M_{1/2-}(L)=\chi_{3+\Delta_{\rm Lin.}}\Lambda_{\rm ALHW}; \qquad {\rm (nucleons\; and \; deltas)}.\label{M_{-;1/2}ALHW}
\end{equation}
For spin-$3/2$ and positive parity, we have
\begin{equation}
    M_{3/2+}(L)=\chi_{3+\Delta_{\rm Lin.}}\Lambda_{\rm ALHW}; \qquad {\rm (nucleons\; and \; deltas)},\label{M_{+;3/2}ALHW}
\end{equation}
and for negative parity,
\begin{equation}
    M_{3/2-}(L)=\chi_{4+\Delta_{\rm Lin.}}\Lambda_{\rm ALHW}; \qquad {\rm (nucleons)}.\label{M_{-;3/2}ALHW}
\end{equation}

\begin{figure}[ht!]
\vskip0.5cm 
\centering
\includegraphics[width=12cm]{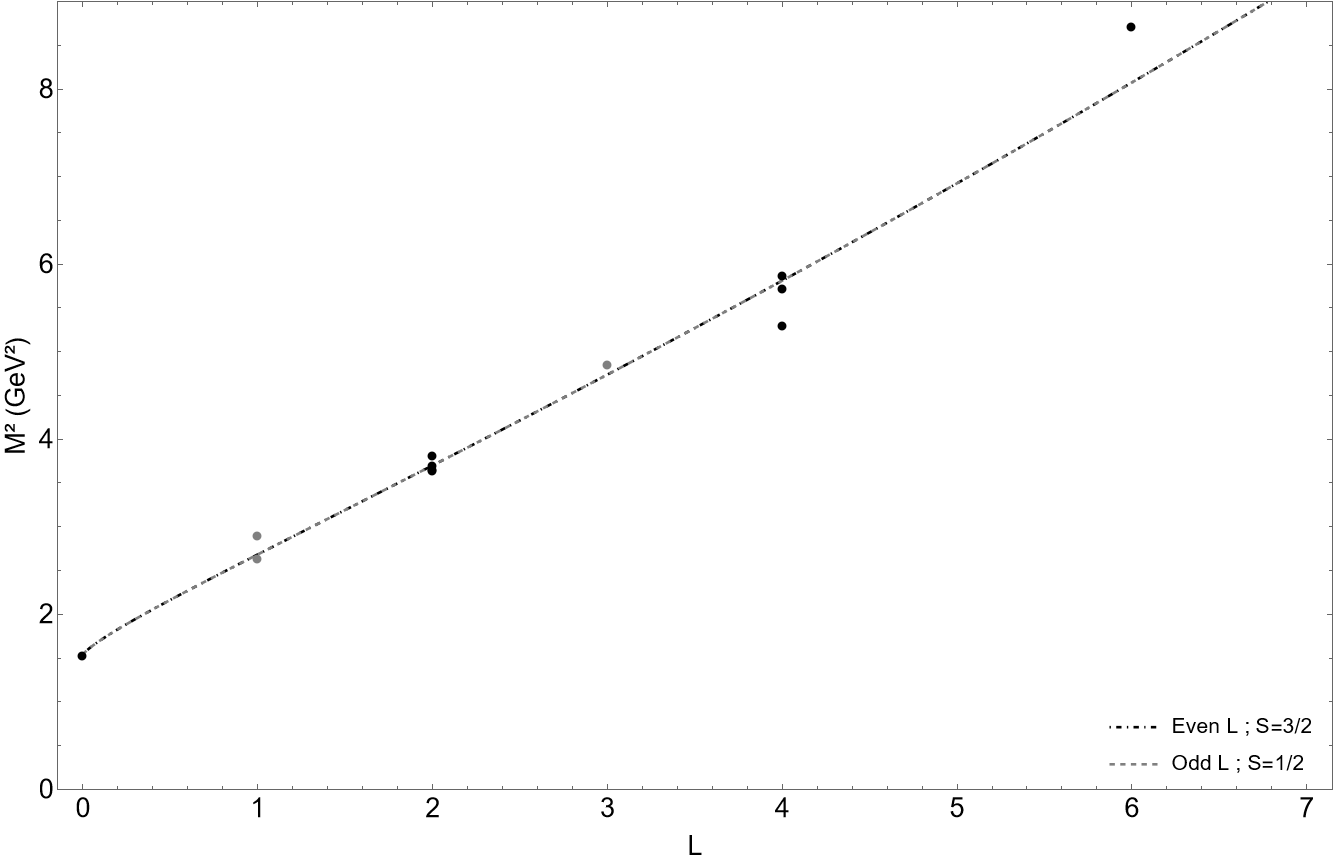}\quad
\includegraphics[width=12cm]{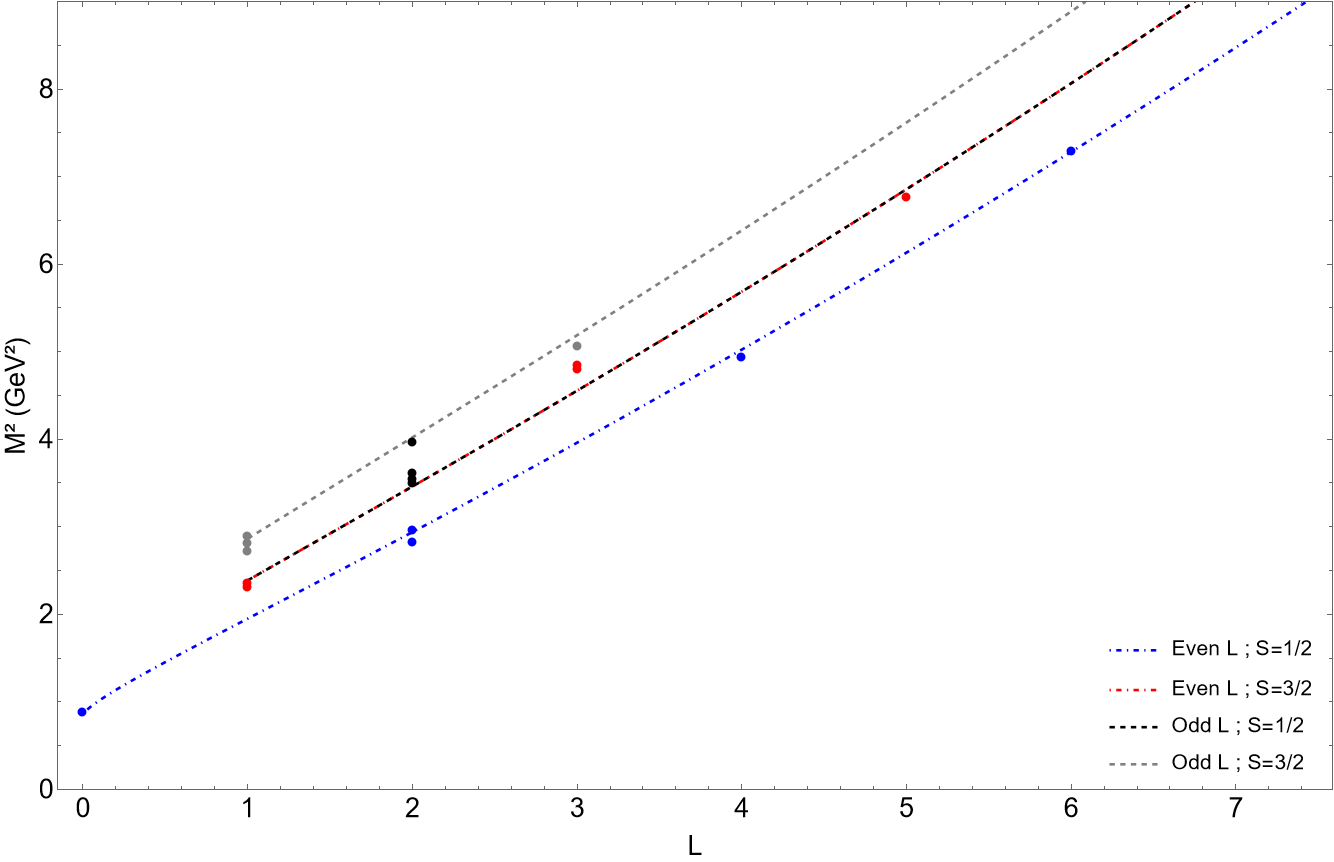}
\caption{Regge trajectories ($L\times M^2$) for baryons, with masses $M$ expressed in GeV, from the anomalous linear hard wall model, obtained from equations (\ref{M_{+;1/2}ALHW})-(\ref{M_{-;3/2}ALHW}), respectively with colors  blue (dot-dashed), black (dashed), red (dot-dashed), and grey (dashed). Note that the trajectories corresponding to Eqs. \eqref{M_{-;1/2}ALHW} (black) and \eqref{M_{+;3/2}ALHW} (red) coincide. {\sl Upper panel:} Delta resonances; {\sl Lower panel:}  Nucleon resonances. In both cases, we also present the points corresponding to the experimental values obtained from the PDG. Each point of a given color corresponds to the trajectory of the same color. }
\label{figALHW}
\end{figure}

Minimizing the function $Q$, as in the previous section, to determine the values of the parameters $a$ and $b$, and with an exponent around $c\approx 0.75$. Figure \ref{figALHW} shows the results obtained for the trajectories, and Table \ref{tabALHW} shows the values of the parameters obtained in this way.  

\vskip 0.5cm
\begin{table}[h!]
\begin{tabular}{|c|c|c|c|c|c|}
\hline
\hline 
Family & $\Lambda$ (MeV) &\, $a$ \, & \, $b$ \, & \, $c$ \,   & Q/n  \%
\\ \hline
Deltas  & 113 &\, 3.12 \, & \, 3.90 \, & \, 0.79 \,  &  (7.54/12)\% = 0.628\%\\
\hline
 Nucleons & 132 & 3.06 & 1.52 & 0.76   & (9.12/18)\% = 0.507\%\\
 \hline\hline
\end{tabular}
\caption{\label{tabALHW} Values of the parameters for the anomalous linear hard wall model in equation (\ref{ALHW}) that minimizes the $Q$ function, Eq. \eqref{qfunction}, for delta and nucleon resonances. The masses obtained in this model are presented in Tables \ref{ALHWDelta} and \ref{ALHWN}, and the Regge trajectories in Fig. \ref{figALHW}.}
\end{table}

\begin{table}[h!]
\centering
\begin{tabular}{|c|c|c|c|c|c|}
\hline\hline
\;$L$ \;& \;$P$ \;& \;$S$ \;& \;PDG \; & \;ALHW\;  & $\delta_{\rm ALHW}$  \\
\hline
0 & + & $\tfrac{3}{2}$ & 1232 & 1240 & 0.64 \\
1 & $-$ & $\tfrac{1}{2}$ & 1620 & 1638 & 1.13 \\
1 & $-$ & $\tfrac{1}{2}$ & 1700 & 1638 & 3.63 \\
2 & + & $\tfrac{3}{2}$ & 1910 & 1924 & 0.72 \\
2 & + & $\tfrac{3}{2}$ & 1920 & 1924 & 0.19 \\
2 & + & $\tfrac{3}{2}$ & 1905 & 1924 & 0.98 \\
2 & + & $\tfrac{3}{2}$ & 1950 & 1924 & 1.35 \\
3 & $-$ & $\tfrac{1}{2}$ & 2200 & 2177 & 1.06 \\
4 & + & $\tfrac{3}{2}$ & 2390 & 2411 & 0.86 \\
4 & + & $\tfrac{3}{2}$ & 2300 & 2411 & 4.81 \\
4 & + & $\tfrac{3}{2}$ & 2420 & 2411 & 0.39 \\
6 & + & $\tfrac{3}{2}$ & 2950 & 2842 & 3.67 \\
\hline\hline
\end{tabular}
\caption{\label{ALHWDelta}  Delta resonance  states characterized by their orbital angular momentum $L$, parity $P$, and spin $S$, with their corresponding masses (in MeV) from PDG, and from the ALHW model, equations (\ref{M_{+;1/2}ALHW})-(\ref{M_{-;3/2}ALHW}), with $\Delta_{\rm Lin.}$ given by equation (\ref{ALHW}). For comparison, we also present the percentage deviation between the masses from this model and PDG.}
\end{table}

\begin{table}[h!]
\centering
\begin{tabular}{|c|c|c|c|c|c|}
\hline\hline
\; $L$\; & \;$P$\; & \;$S$\; &\; PDG\; &\; ALHW\;  & $\delta_{\rm ALHW}$  \\
\hline
0 & + & $\tfrac{1}{2}$ & 939 & 926 & 1.43 \\
1 & $-$ & $\tfrac{1}{2}$ & 1520 & 1551 & 2.02 \\
1 & $-$ & $\tfrac{1}{2}$ & 1535 & 1551 & 1.03 \\
1 & $-$ & $\tfrac{3}{2}$ & 1650 & 1700 & 3.05 \\
1 & $-$ & $\tfrac{3}{2}$ & 1700 & 1700 & 0.018 \\
1 & $-$ & $\tfrac{3}{2}$ & 1675 & 1700 & 1.51 \\
2 & + & $\tfrac{1}{2}$ & 1720 & 1718 & 0.088 \\
2 & + & $\tfrac{1}{2}$ & 1680 & 1718 & 2.29 \\
2 & + & $\tfrac{3}{2}$ & 1880 & 1867 & 0.70 \\
2 & + & $\tfrac{3}{2}$ & 1900 & 1867 & 1.75 \\
2 & + & $\tfrac{3}{2}$ & 1870 & 1867 & 0.17 \\
2 & + & $\tfrac{3}{2}$ & 1990 & 1867 & 6.19 \\
3 & $-$ & $\tfrac{1}{2}$ & 2190 & 2142 & 2.21 \\
3 & $-$ & $\tfrac{1}{2}$ & 2220 & 2142 & 3.53 \\
3 & $-$ & $\tfrac{3}{2}$ & 2250 & 2288 & 1.67 \\
4 & + & $\tfrac{1}{2}$ & 2220 & 2247 & 1.23 \\
5 & $-$ & $\tfrac{1}{2}$ & 2600 & 2627 & 1.05 \\
6 & + & $\tfrac{1}{2}$ & 2700 & 2706 & 0.22 \\
\hline\hline
\end{tabular}
\caption{\label{ALHWN} Nucleon resonance  states characterized by their orbital angular momentum $L$, parity $P$, and spin $S$, with their corresponding masses (in MeV) from PDG, and from the ALHW model, equations (\ref{M_{+;1/2}ALHW})-(\ref{M_{-;3/2}ALHW}), with $\Delta_{\rm Lin.}$ given by equation (\ref{ALHW}). For comparison, we also present the percentage deviation between the masses from this model and PDG.}
\end{table}

\FloatBarrier
\section{Discussion and Conclusions}

    In this work, anomalous modifications of the original Hard Wall model were proposed to describe light baryons and obtain their spectra. We also presented the original Hard Wall model using different canonical dimensions for spin states $S=1/2$ and $S=3/2$. The main objective of such changes was to obtain better out puts in comparison with previous results from the Hard Wall model for baryons.

    Despite the fact that we obtain better results only by changing the canonical dimension, the HW model still presents non linear Regge trajectories and that was one of the reasons for considering the AHW models.

    Inspired by the semiclassical approximation of the AdS/CFT correspondence for the anomalous dimension, given by Eq.  (\ref{anomalousdimension}), we proposed two models for our anomalous dimension. The first depends only on the orbital angular momentum, given by equation (\ref{deltaanom1}). This model allowed us to obtain approximate linear Regge trajectories and better masses for baryons if compared with PDG data. The second depends also on spin, equation (\ref{deltaanom2}). This model presents trajectories that are also more linear than the original hard wall, while the obtained masses are not as good as the first anomalous model. This comparison can be made more precise by looking at the values of $Q/n$ in Table \ref{tabmodelinfo}, where the values are almost 50\% \ better in AHW$_1$ compared to AHW$_2$. If we compare the Regge trajectories of these models, in the first case the masses from PDG are in better agreement with the trajectories on which they should rely.

    In addition, the linear realization of the anomalous dimension discussed in Section VI provides further insight into the asymptotic structure of the spectrum. In contrast to the purely logarithmic implementations, the linear correction improves the large-spin behavior of the Regge trajectories, yielding a more uniform spacing among highly excited states while preserving the overall consistency with the experimental data. Although this modification does not drastically alter the low-lying spectrum, it suggests that the functional form of the anomalous contribution plays a nontrivial role in controlling the asymptotic regime of the model. These results indicate that refining the scaling dimension beyond canonical values — particularly through simple power law  deformations — may offer a promising direction for achieving a better holographic description of baryon Regge trajectories within the Hard Wall framework.

The possibility of a deeper physical motivation in the proposed anomalous linear model should be investigated in future work. Although it may not seem simple to demonstrate its phenomenological validity from first principles, we can perform an analysis, possibly studying objects with more valence quarks, such as tetraquarks and pentaquarks.

\begin{acknowledgments} 
RACS is supported by Coordenação de Aperfeiçoamento de Pessoal de Nível Superior (CAPES). HBF is partially supported by Conselho Nacional de Desenvolvimento Cient\'{\i}fico e Tecnol\'{o}gico (CNPq) under grant  310346/2023-1, and Fundação Carlos Chagas Filho de Amparo à Pesquisa do Estado do Rio de Janeiro (FAPERJ) under grant E-26/204.095/2024.
\end{acknowledgments}

\end{document}